**MassChroQ : A versatile tool for mass spectrometry quantification**

Benoit Valot[1], Olivier Langella[2], Edlira Nano[1], Michel Zivy[2].

[1]*INRA, Plateforme d'Analyse Protéomique de Paris Sud-Ouest, UMR 0320/UMR 8120 de Génétique Végétale, Gif sur Yvette, France.*
[2]*CNRS, PAPPSO, Plateforme d'Analyse Protéomique de Paris Sud-Ouest, UMR 0320/UMR 8120 de Génétique Végétale, Gif sur Yvette, France.*

Correspondence :

Dr Benoît Valot, PAPPSO, UMR de Génétique Végétale, INRA/Université Paris-Sud/ CNRS/AgroParisTech, Ferme du Moulon, F-91190 Gif-sur-Yvette, France.

Email : valot@moulon.inra.fr

Fax : +33 1 69 33 23 40

Abbreviations :

MassChroQ, Mass Chromatogram Quantification; HR, High Resolution; LR, Low Resolution; RT, Retention Time;  XIC, eXtracted Ion Chromatogram

Number of words : 2773

**Abstract**

Recently, many software tools have been developed to perform quantification in LC-MS analyses. However, most of them are specific to either a quantification strategy (e.g. label-free or isotopic labelling) or a mass-spectrometry system (e.g. high or low resolution).

In this context, we have developed MassChroQ, a versatile software that performs LC-MS data alignment and peptide quantification by peak area integration on extracted ion chromatograms.

MassChroQ is suitable for quantification with or without labelling and is not limited to high resolution systems. Peptides of interest (for example all the identified peptides) can be determined automatically or manually by providing targeted *m/z* and retention time values. It can handle large experiments that include protein or peptide fractionation (as SDS-PAGE, 2D-LC). It is fully configurable. Every processing step is traceable, the produced data are in open standard format and its modularity allows easy integration into proteomic pipelines. The output results are ready for use in statistical analyses.

Evaluation of MassChroQ on complex label-free data obtained from low and high resolution mass spectrometers showed low CVs for technical reproducibility (1.4%) and high coefficients of correlation to protein quantity (0.98).

MassChroQ is freely available under the GNU General Public Licence v3.0 at http://pappso.inra.fr/bioinfo/masschroq/.

In the last years, the continuous improvement of quantitative mass spectrometry methods has opened new perspectives in proteomics. The amount and complexity of the data to be processed has grown, evidencing the need for new automatic computational tools. While spectral counting has been used for semi-quantitative analysis, quantitative experiments are mostly based on quantification of the MS signal. Several tools using various algorithms have been developed (e.g [1-3]). These tools handle the different steps of the analysis : signal denoising, peak detection, peak area measurement, de-isotoping, LC-MS runs alignment, etc. However, most of them deal only with a specific problem or type of data, and are for example restricted to high-resolution (HR) spectrometers, to isotope labelling or to label-free quantification. In addition, they often present platform specificities, proprietary data format dependencies and do not allow integration in proteomics pipelines like TPP [4] or TOPP [5].

We have developed MassChroQ (which stands for Mass Chromatogram Quantification) with the aim of being as experiment-independent as possible, while being able to take into account complex experimental designs. MassChroQ processes quantification data from their rough state to a form ready to be used by statistical software. It is fully configurable and every step of the analysis is traceable. MassChroQ allows the user to: i) process data obtained from spectrometers with various levels of resolution; ii) analyse label-free as well as isotopic labelling experiments; iii) analyse experiments in which samples were fractionated prior to LC-MS analysis (as in SDS-PAGE, SCX, etc.); iv) time-efficiently process a large number of samples.

Low-resolution (LR) instruments (e.g. LTQ ion traps) can provide valuable quantitative data from samples of low or medium complexity. In order to be able to quantify data obtained from LR as well as from HR instruments (e.g. Orbitrap), we chose a quantification method based on eXtracted Ion Chromatograms (XIC) rather than on feature detection on the virtual 2D image (e.g algorithms derived from 2D gels analysis or "peak picking" [6-10]). Indeed, the latter needs high resolution in MS mode to identify isotopic profiles. By contrast, quantification based on XICs is obtained by extracting the intensity corresponding to the *m/z* of the selected peptides along the LC-MS run, and by integrating the peak area at their retention time (RT). This strategy can be used with LR as well as HR mass spectrometers by adapting the window size of XIC extraction. It can be used with label-free (e.g. [11,12] as well as isotopic methods (e.g. SILAC, ICAT, $N_{15,}$ [13-15]).

The main features of MassChroQ are :

i) *Determination of peptides to be quantified.* If an experiment includes MS/MS acquisition for identification, the identified peptides and protein descriptions can be provided to MassChroQ. He will then automatically quantify them in all samples, including those where the peptides were not identified. Peptides can also be specified by providing a list of *m/z* or *m/z*-RT values. If isotopic labelling was performed, the different labels can be described by specifying the modified sites (e.g. amino acids, peptide N- or C-terminal) and the mass shifts.

ii) *XIC extraction, peak detection and quantification.* XICs of peptides of interest are extracted from the original data file. Filters are used to correct baselines or to remove artefactual spikes (Fig S1). XICs are then smoothed with an average filter before performing a closing and an opening mathematical morphology operation with a small flat structuring element [16]. The closing operation eliminates thin valleys and conserves the intensity of local maxima, while the opening operation eliminates thin peaks (i.e. remaining spikes) and conserves the intensity of local minima. Hence, detection of peak positions is performed on the closed profile, and the opened profile is used to eliminate remaining spikes (Fig. S2). The peak boundaries are searched on the closed profile, and the peak area (i.e. the quantification value) is computed on the unaltered XIC, by integrating the intensity between these boundaries.

iii) *Alignment and peak matching.* As distortions can occur between LC-MS runs, MS RTs must be aligned before matching. Two alignment methods are proposed in MassChroQ: the OBI-Warp (Ordered Bijective Interpolated Warping) alignment method [17] which is based on MS data only, and an in-house MS/MS alignment method. The latter uses common MS/MS identifications as landmarks to evaluate time deviation ($\Delta_{MS}$) along the chromatography. More precisely, the $\Delta_{MS/MS}$ difference between the MS/MS RT in the run to align and the MS/MS RT in the run chosen as a reference is computed for each common peptide. Then, for each MS RT of the run to align, $\Delta_{MS}$ is evaluated by linear interpolation between the $\Delta_{MS/MS}$ values of its two closest surrounding MS/MS RTs. Both $\Delta_{MS/MS}$ and $\Delta_{MS}$ data are smoothed before use (with average and median filters) to eliminate low-scale RT heterogeneities. After alignment, peak matching is performed as follows: the quantitative value of a peak is assigned to an identified peptide if and only if the MS/MS RT of this peptide is within the boundaries of this peak. Of course, only similar LC-MS runs should be aligned. For example, if samples were fractionated by SCX, only LC-MS runs from the same SCX fraction should be aligned. To take this into account the user defines groups of LC-MS runs that can be compared to each other. Alignment and peak matching will be performed only within these groups. If necessary, alignment and quantification methods can be specifically defined for each group.

To evaluate MassChroQ performances, we prepared 6 samples made each of 700 ng of the same total protein digest of *Saccharomyces cerevisiae,* spiked with 6 different amounts of BSA digest (4.5, 15, 45, 105, 450 and 1500 fmol). These samples were analysed with an LR and an HR system (respectively a Thermo-Fisher LTQ XL coupled to an Eksigent 2D-ultra-nanoLC, and a Thermo-Fisher Orbitrap Discovery coupled to a Dionex U3000 nanoLC; see supplement materials). All runs included MS and MS/MS acquisition. Two groups of runs were defined in MassChroQ to separate the six LR runs from the six HR runs. The alignment was performed by using the MS/MS alignment method (Fig. 1A). Since spectrometers do not trigger MS/MS at the exact RT of peptide peaks, MS/MS RTs showed a non negligible dispersion. However, data points were numerous enough to allow the computation of the tendency deviation curve along the reference LC, that was used by the alignment algorithm to correct RTs . The standard deviation of peptide RT was clearly reduced by the alignment in both LR and HR systems (Fig. 1B and 1C), showing its efficiency. Although LC-MS runs showed small deviations before alignment, the alignment significantly impacted data quality affecting the matching of 5% of the peaks (data not shown).

XIC extraction was performed with an m/z window of 0.3 Th for LR data and 10 ppm for HR data. All identified peptides were selected for quantification. Combining all LR and HR LC-MS/MS runs, 5831 different peptide sequences allowed the identification of 556 proteins (with a false discovery rate of 0.3%), distributed in 492 groups of proteins sharing at least one peptide. A total of 5936 and 2467 XICs were extracted from respectively LR and HR LC-MS runs. Almost all detected peptides were found reproducible (i.e. detected in at least five of the six replicates) in the HR system (97%), against 67% in the LR system (Fig. 2A). Peptide reproducibility was clearly correlated to peptide intensity in LR data (Fig. S3), most probably due to noisy XICs. Altogether, 418 of the 492 identified proteins were represented by at least one reproducible peptide.

After normalization and $\log_{10}$-transformation (see supplement materials), the mean coefficient of variation of peptide quantitative values was 1.31% in HR and 1.40% in LR data (Fig. 2B, 2C). This small technical variation is similar to other reported data (see [1]) and attests the accuracy of the detection/quantification process. Moreover, a correlation of 0.89 between the mean intensity of peptides common to LR and HR data (1179 peptides, Fig. 2D) showed that the quantification process extracted similar results from both systems, despite a high sample complexity not favourable to LR analysis. The few high coefficients observed for abundant peptides in the HR data were mostly due to a poor determination of the ends of smearing peaks.

Twenty-five and fourteen BSA peptides were quantified in at least three samples in respectively LR and HR systems. All HR peptide intensities except one were highly correlated and linearly related to injected BSA quantities with a mean coefficient of correlation greater than 0.98 on three orders of magnitude. This exception was due to a single datapoint (Fig. 3A). Nineteen of the twenty-five LR peptides responded linearly to BSA quantities with a mean coefficient of correlation higher than 0.98 on two orders of magnitude (Fig. 3B). The lower correlation observed for the six remaining peptides was mainly due to miss-assignments at low BSA quantities (<45 fmol): the BSA peptide peak was contaminated by a peak of the yeast digest of similar *m/z* and RT values (Fig. S4). Thus, quantification performances were lower with the LR than with the HR system, mainly because of mismatches caused by the high complexity of the yeast lysate. This confirms that accurate measurements can be expected with LR systems only when analysing peptide samples of lower complexity. Nevertheless, the observed correlations between peptide intensity and protein quantity were globally similar to those obtained by other software [4,5,12].

MassChroQ is written in C++ with Qt and runs both on Linux and Windows platforms. It is a command-line standalone program and it comes with a library for integration in proteomics pipelines.

MassChroQ is fully configurable via an XML input file (in masschroqML format) where the user indicates the chosen processing steps, parameters and data files to analyse (see example on Fig. S5). This file can be automatically generated by any XML editor by using the provided schema, or manually by using a text editor. Parameters of XIC creation, filtering and detection, which depend on the type, precision and noise level of the spectrometer, can all be configured in the masschroqML file. Templates for several experiment scenarios are provided in the documentation.

LC-MS data input files can be in mzXML [18] or mzML format [19]. If X!Tandem [20] is used for protein identification, a complete masschroqML input file containing identified peptides and protein descriptions can be automatically generated via our X!Tandem pipeline tool (http://pappso.inra.fr/bioinfo/xtandempipeline/). If another identification engine is used, identified peptides to be quantified can be provided to MassChroQ via TSV or CSV text files (Tab or Comma Separated Values). MassChroQ results can be exported in TSV, gnumeric spreadsheet or masschroqML XML format. TSV and spreadsheet formats allow direct import of data to statistical software and the XML format allows their upload in proteomics databases like PROTICdb [21]. XICs can also be exported for visualization.

Computation time depends on data size and on the number of extracted XICs. In the present study, the processing of the twelve LC-MS runs (6GB) where more than 5000 different peptide XICs were extracted took 1 hour with a 2.93 GHz CPU on a Linux platform. Most of that time was spent analysing non-centroid data from the LR system.

In conclusion, we showed that MassChroQ efficiently aligns and quantifies LR and HR LC-MS data. Low coefficients of variation and high coefficients of correlation to protein quantity attested the quality of the quantification measurements. MassChroQ is currently being successfully used in our laboratory on both isotopic and label-free large experiments (data not shown). Its very modular structure facilitates implementation of new algorithms and integration in other pipelines. Future developments will focus on handling SRM data, developing a graphical user interface for parameter adjustment with XIC visualization and computation time optimisation. This program is licensed under the GNU General Public License v3.0 (http://www.gnu.org/licenses/gpl.html). The source code is available at https://sourcesup.cru.fr/projects/masschroq. Compiled binaries for Linux and Windows platforms and documentation (including data and results of this test set) can be found at http://pappso.inra.fr/bioinfo/masschroq/.

**Acknowledgements**

This work was partially supported by "Infrastructures en Biologie Santé et Agronomie (IbiSA)" (E N salary). The authors want to thank Mélisande Blein for the yeast extracts and SourceSup for their subversion repository support.

**Conflict of interest statement**

The authors have declared no conflicts of interest.

2081.

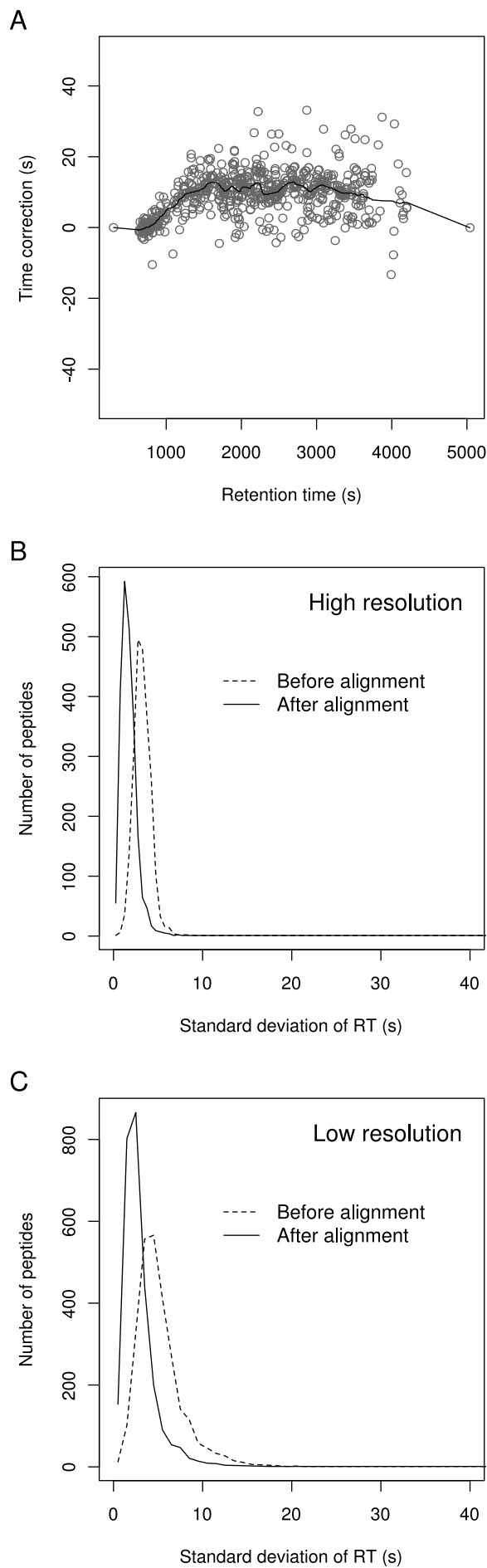

**Figure 1.** Evaluation of alignment. (A) Example of RT correction along an HR LC-MS run. Points correspond to peptides identified by MS/MS in both the reference LC-MS run and the LC-MS run being aligned. The line is the computed time deviation used for alignment of MS data. (B, C) Standard deviation of peptide RT before and after alignment respectively in HR and LR experiments.

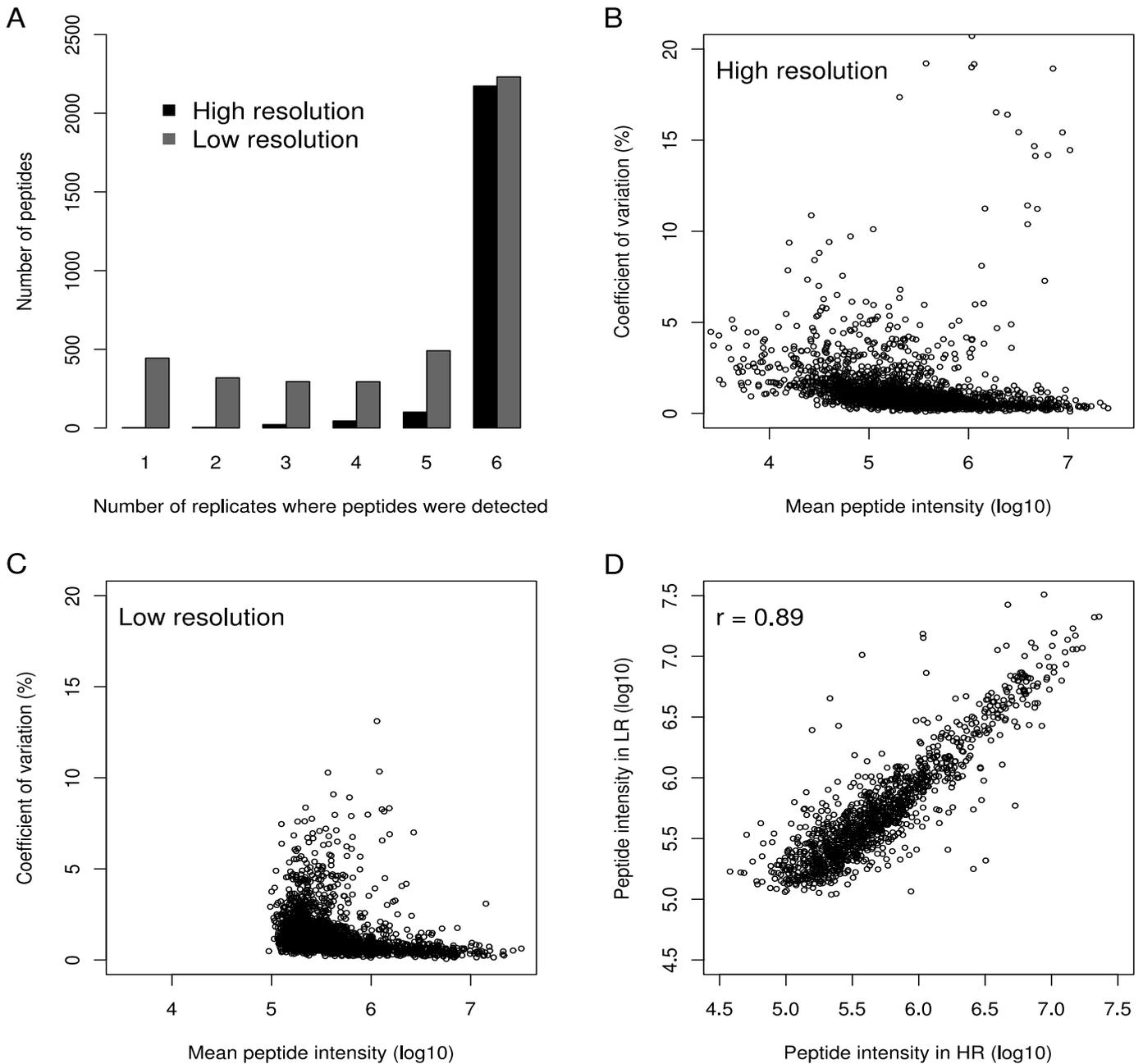

**Figure 2.** Evaluation of reproducibility. (A) Number of peptides detected in 1 to 6 replicates in LR and HR LC-MS experiments. (B, C) Influence of mean peptide intensity on peptide coefficient of variation (all reproducible peptides) in HR and LR systems. (D) Correlation between HR and LR mean values of peptides identified in both systems (1179 peptides, r = 0.89).

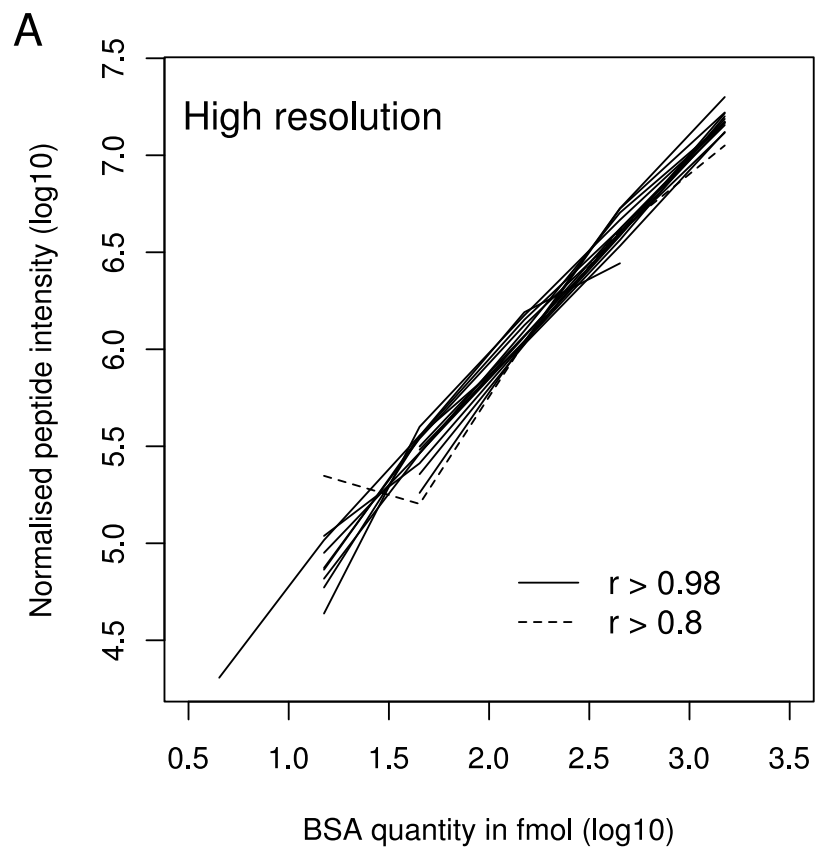

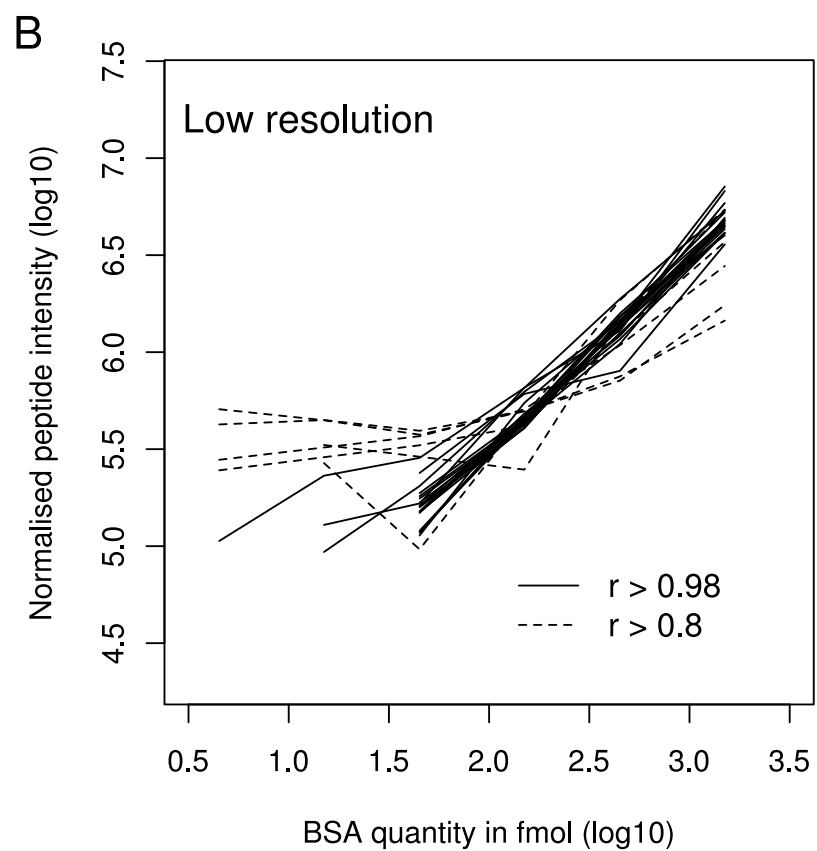

**Figure 3.** Linear relation between BSA peptide intensity and BSA quantity. (A) and (B) : respectively HR (25 peptides) and LR (14 peptides) experiments. Each line corresponds to a different peptide. Only peptides detected in at least 3 samples are shown. Dashed line: peptides showing a correlation to BSA quantity lower than 0.98.

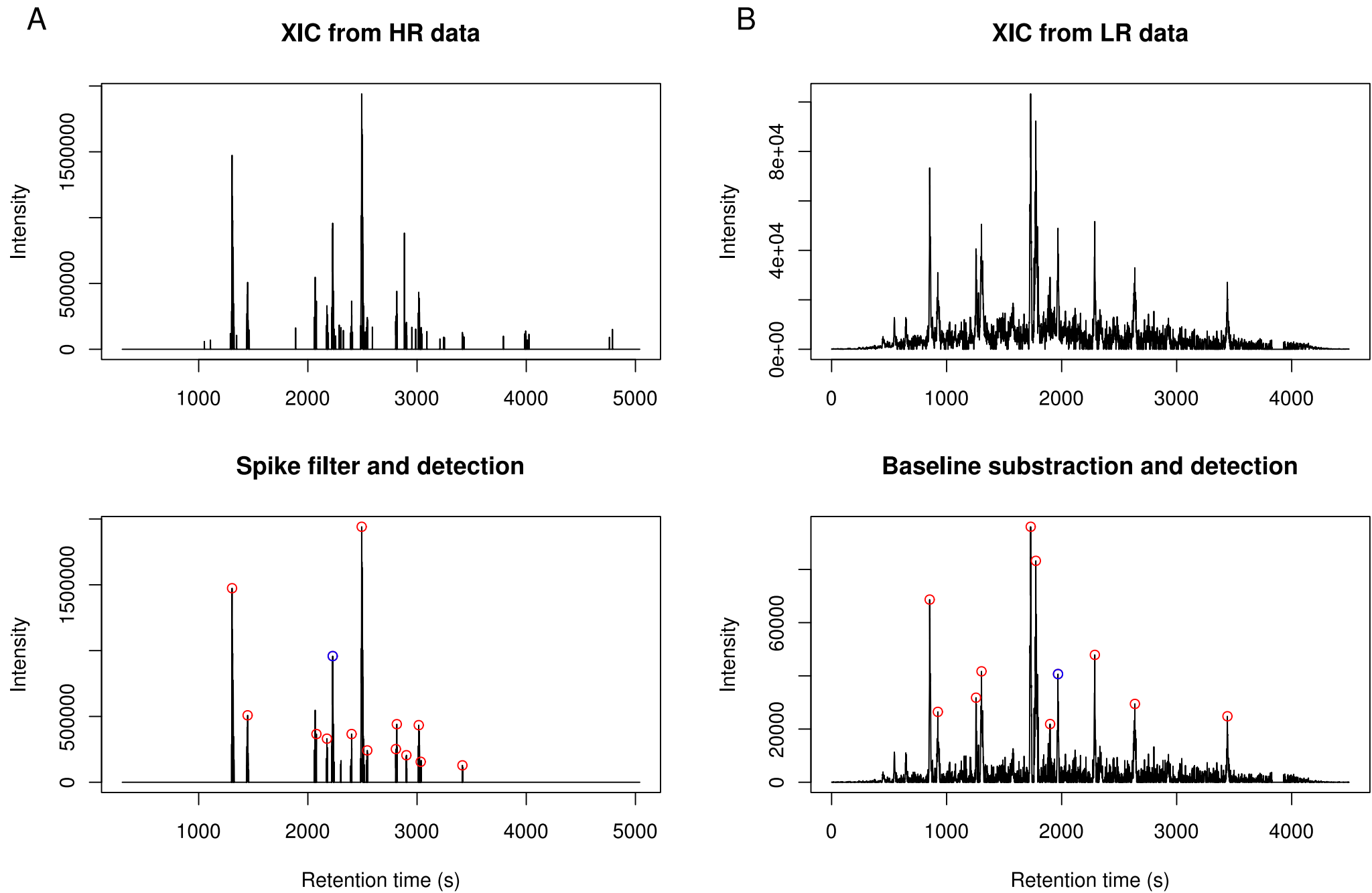

**Figure S1.** XIC creation, filtering and peak detection: example for peptide LVNELTEFAK (45 fmol of BSA digest spiked in yeast total digest) in LC-MS runs from LR (A) and HR (B) experiments. Filtering involves baseline correction for LR experiments and spike removing for HR ones. Red circles: detected peaks; blue circles: detected peak matching the peptide's RT.

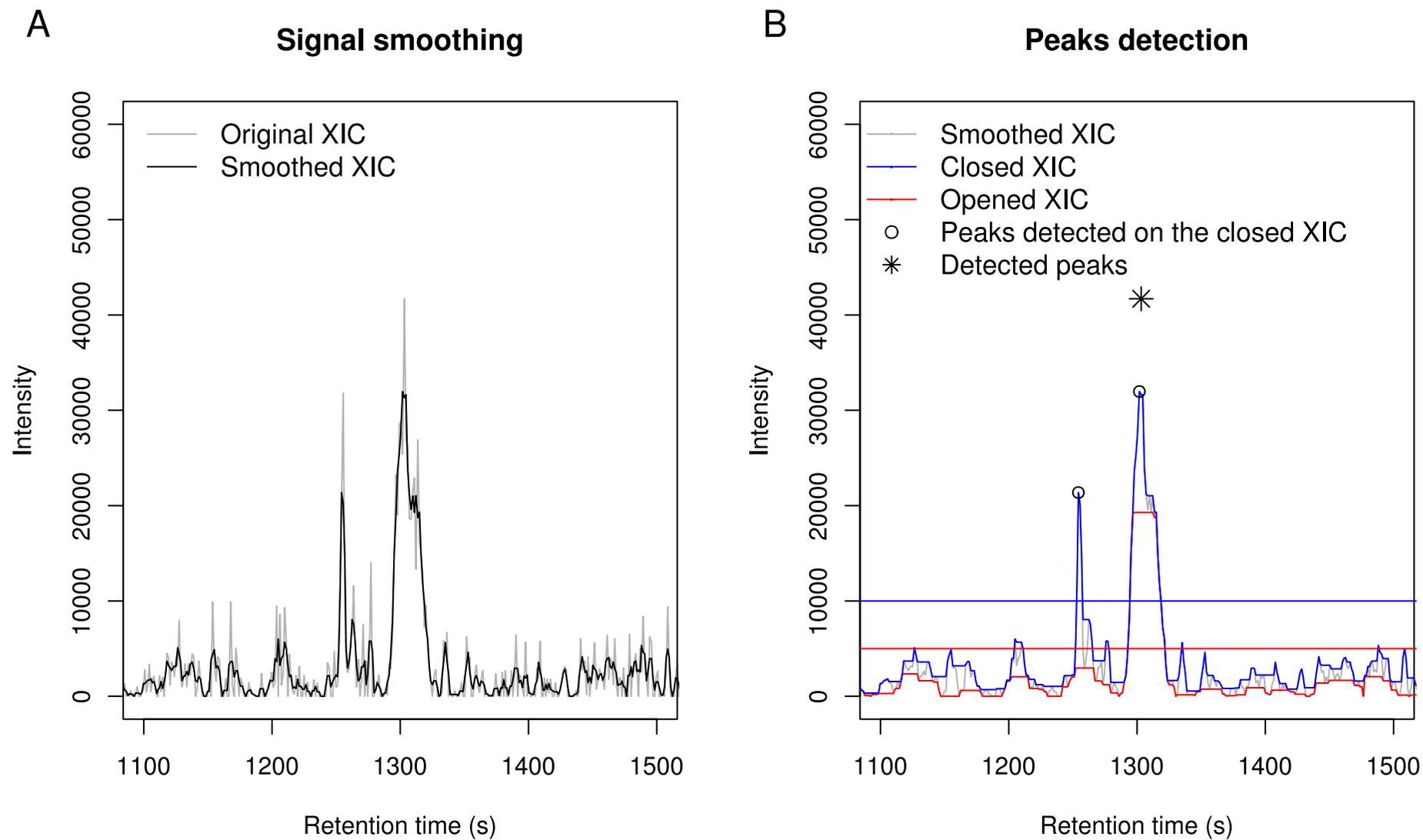

**Figure S2.** Signal treatment for peak detection. (A) The original XIC contains high frequency noise that is partly eliminated by smoothing (average and/ or median filters). (B) Morphological closing (blue profile) and opening (red profile) by a small flat structuring element are performed on the smoothed XIC (gray profile). The closing operation eliminates many noisy peaks by filling small valleys and preserves the actual position of the remaining peaks. Hence peaks are detected on this profile if they are greater than a threshold (blue line). Only peaks that are thick compared to the structuring element stay high in the opened profile. Then, to avoid detection of thin artifactual spikes, peaks detected on the closed profile are filtered according to the intensity at the same position in the opened profile : intensity in the opened profile must be greater than a second threshold (red line).

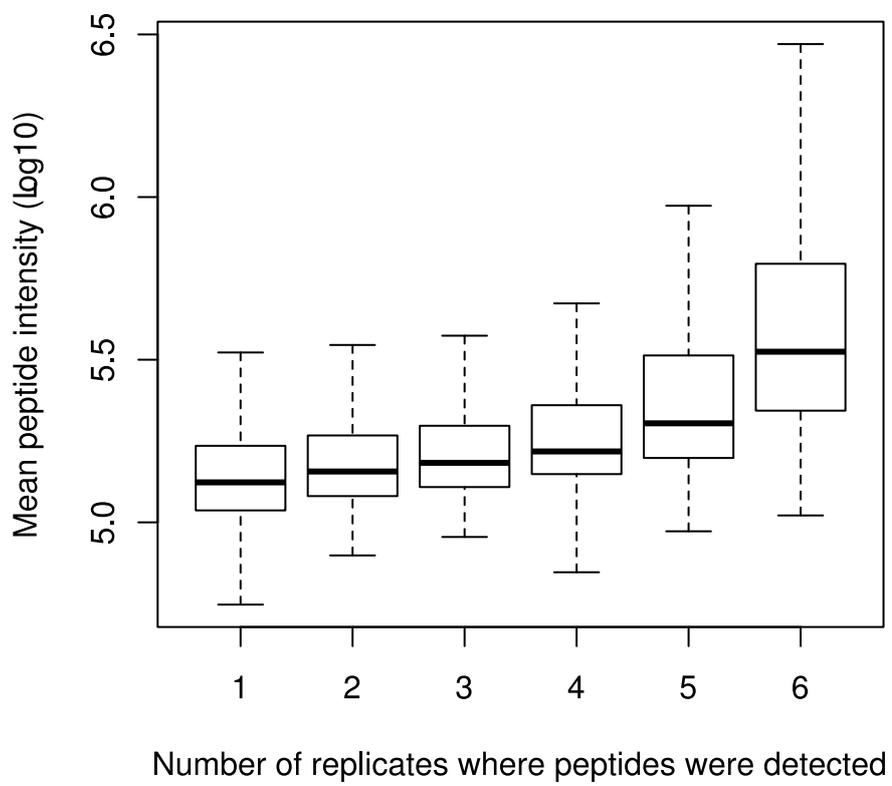

**Figure S3.** Relation between peptide intensity and detection reproducibility in the LR experiment.

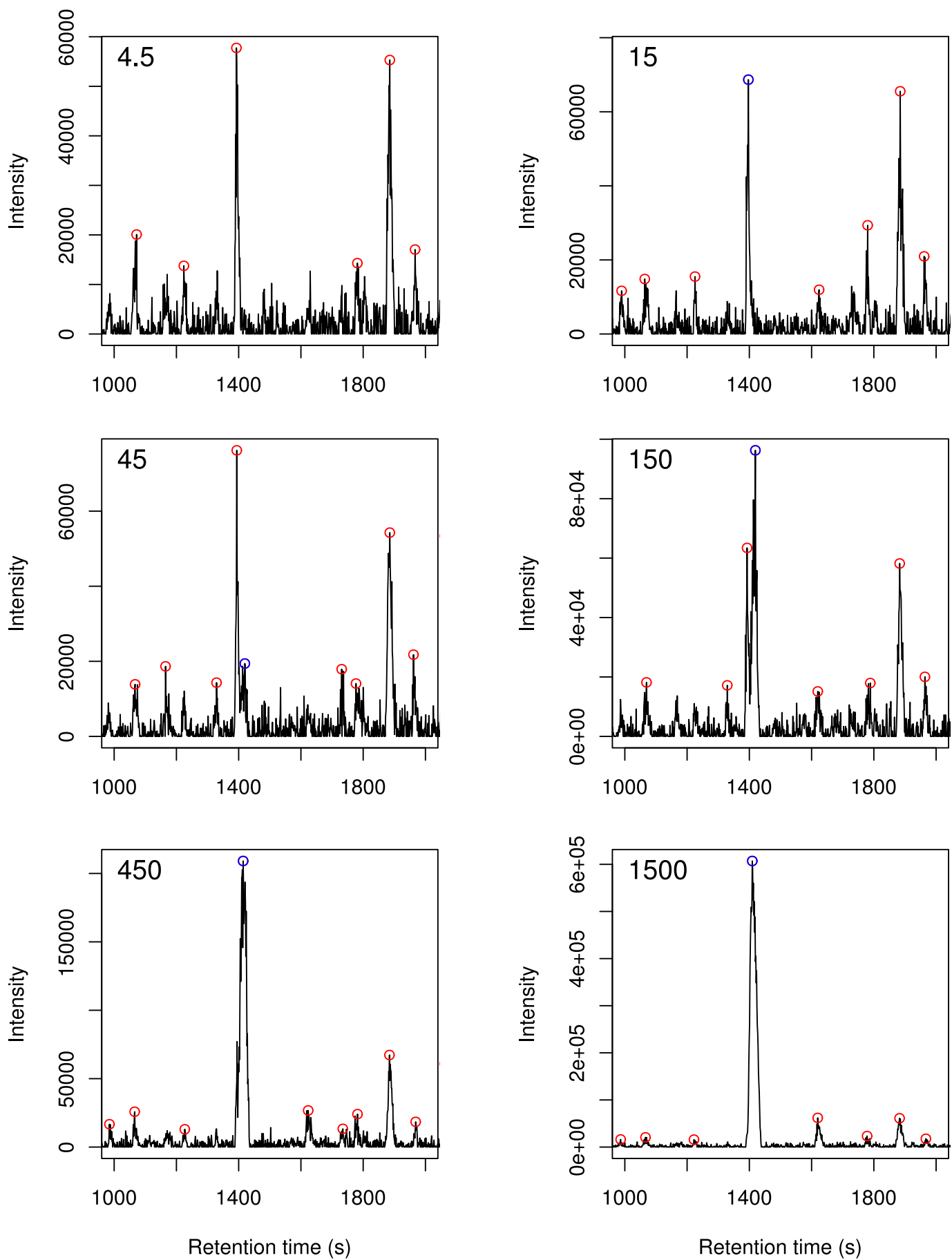

**Figure S4.** Peak assignment error in LR data analysis: example of the BSA peptide HLVDEPQNLIK.
The BSA digest quantity is indicated on the top-left corner of each graph. Red circles: detected peaks. Blue
circles: Peaks matching the peptide's RT. A yeast peak at RT and m/z values very close to those of
HLVDEPQNLIK, induced a mismatch at 15 fmol. In addition, peaks assigned to HLVDEPQNLIK in other
samples partly contained the intensity that should have been assigned to the yeast peptides.

```xml
<!--Example of MassChroQ processing file-->
<?xml version="1.0" encoding="UTF-8" standalone="no"?>
<masschroq>

<!-- List of LC-MS run files in open format : mzXML or mzML -->
<rawdata>
<data_file id="samp0" format="mzxml" path="Labelling-light-heavy.mzXML" type="centroid"/>
<data_file id="samp1" format="mzxml" path="Label-free-samp1.mzXML" type="centroid"/>
<data_file id="samp2" format="mzxml" path="Label-free-samp2.mzXML" type="centroid"/>
<data_file id="samp3" format="mzxml" path="MS-data-without_identification-samp1.mzXML"
        type="profile"/>
<data_file id="samp4" format="mzxml" path="MS-data-without_identification-samp2.mzXML"
        type="profile"/>
</rawdata>
<groups>

<!-- Grouping of LC-MS runs. Within a group:
    - all LC-MS runs are aligned (with the same alignment method);
    - peptides observed in at least one LC-MS run are quantified in all LC-MS runs
    of this group (using the same quantification method) -->
<group data_ids="samp0" id="G1"/>
<group data_ids="samp1 samp2" id="G2"/>
<group data_ids="samp3 samp4" id="G3"/>
</groups>

<!-- The peptide features list can be defined in two ways :
    in separate spreadsheets containing identified peptides for each LC-MS run as follows :-->
<peptide_files_list>
<peptide_file data="samp0" path="labelling_peptide_list.txt"/>
</peptide_files_list>
<!-- directly in this file with proteins/peptides list as follows : -->
<protein_list>
<protein desc="conta|P02769|ALBU_BOVIN SERUM ALBUMIN PRECURSOR." id="P1.1"/>
</protein_list>
<peptide_list>
<peptide id="pep0" mh="1463.626" mods="114.08" prot_ids="P1.1" seq="TCVADESHAGCEK">
<observed_in data="samp2" scan="755" z="2"/>
<observed_in data="samp3" scan="798" z="2"/>
</peptide>
<peptide id="pep1" mh="1103.461" mods="57.04" prot_ids="P1.1" seq="ADESHAGCEK">
<observed_in data="samp2" scan="663" z="2"/>
</peptide>
</peptide_list>

<!-- Definition of different labels for isotopic experiments.
    Example with a dimethylation of primary amine -->
<isotope_label_list>
<isotope_label id="light">
<mod at="Nter" value="28.0"/>
<mod at="K" value="28.0"/>
</isotope_label>
<isotope_label id="heavy">
<mod at="Nter" value="32.0"/>
        <mod at="K" value="32.0"/>
</isotope_label>
</isotope_label_list>

<!-- Definition of different alignment methods. Two alignment algorithms
    are implemented : MS/MS alignment and OBI-Warp alignment -->
<alignments>
<alignment_methods>
<alignment_method id="ms2_1">
<ms2>
<ms2_tendency_halfwindow>10</ms2_tendency_halfwindow>
<ms2_smoothing_halfwindow>5</ms2_smoothing_halfwindow>
<ms1_smoothing_halfwindow>0</ms1_smoothing_halfwindow>
</ms2>
</alignment_method>
```

```xml
<alignment_method id="obiwarp1">
<obiwarp>
<lmat_precision>1</lmat_precision>
<mz_start>500</mz_start>
<mz_stop>1200</mz_stop>
</obiwarp>
</alignment_method>
</alignment_methods>
<!-- Perform alignment on each group using the above defined methods.
     A reference LC-MS run should be defined for each group. All other runs of
     the group will be aligned towards this reference run -->
<align group_id="G2" method_id="ms2_1" reference_data_id="samp1"/>
<align group_id="G3" method_id="obiwarp1" reference_data_id="samp3"/>
</alignments>

<!-- Definition of different quantification methods and parameters for
     XIC creation, XIC filtering and peak detection -->
<quantification_methods>
<quantification_method id="quanti1">
<!-- XIC creation on mz or ppm range using TIC (sum) or basepeak (max) -->
<xic_extraction xic_type="sum">
<ppm_range max="10" min="10"/>
</xic_extraction>
<!-- XIC filtering with spike removing, baseline correction or smoothing -->
<xic_filters>
<anti_spike half="5"/>
<background half_mediane="5" half_min_max="15"/>
<smoothing half="3"/>
</xic_filters>
<!-- XIC detection with threshold -->
<peak_detection>
<detection_zivy>
<mean_filter_half_edge>1</mean_filter_half_edge>
<minmax_half_edge>3</minmax_half_edge>
<maxmin_half_edge>2</maxmin_half_edge>
<detection_threshold_on_max>5000</detection_threshold_on_max>
<detection_threshold_on_min>3000</detection_threshold_on_min>
</detection_zivy>
</peak_detection>
</quantification_method>
</quantification_methods>

<!-- Quantification area -->
<quantification>
<!-- Definition of the export files and formats for the quantification results :
     spreadsheet formats (tsv, gnumeric) or xml format -->
<quantification_results>
<quantification_result output_file="results" format ="tsv"/>
</quantification_results>

<!-- Definition of the export files for the XIC traces in spreadsheet (tsv) format :
     one can trace all the XICs and/or a list of given peptides and/or a list of
     given m/z values, and/or a list of m/z-rt values -->
<quantification_traces>
<all_xics_traces output_dir="all_xics_traces" format="tsv"/>
<peptide_traces peptide_ids="pep0 pep1" output_dir="peplist_xics_traces" format="tsv"/>
</quantification_traces>

<!-- For each group, start quantification on : -->
<!-- all the peptides -->
<quantify withingroup="G1" quantification_method_id="quanti1">
<peptides_in_peptide_list mode="real_or_mean" isotope_label_refs="light heavy"/>
</quantify>
<quantify withingroup="G2" quantification_method_id="quanti1" id="q1">
<peptides_in_peptide_list mode="real_or_mean"/>
</quantify>
<!-- and/or on a list of given m/z values -->
<quantify withingroup="G3" quantification_method_id="quanti1" id="q2">
<mz_list>732.317 449.754 552.234 464.251 381.577 569.771 575.256</mz_list>
</quantify>
<!-- Quantification can also be performed on a list of m/z-rt values -->
</quantification>
</masschroq>
```

**Supplementary materials and methods**

**1 Protein extraction and digestion**

The proteins were extracted from yeast cell pellets using a TCA/acetone method: proteins were precipitated in 10% TCA, 0.07% 2-Mercaptoethanol in acetone and the pellet was rinsed in 0.07% 2-Mercaptoethanol in acetone. Precipitated proteins were resuspended in urea 8 M, thiourea 2 M, CHAPS 2%. Protein concentration was determined by using the PlusOne 2-D Quant Kit (GE Healthcare), and 10 µg of proteins were reduced and alkylated with respectively 10 mM of DTT and 55 mM of iodoacetamide. Proteins were diluted 10 times with 50 mM ammonium bicarbonate and digested with 1/50 (w/w) trypsin (Promega) at 37 °C overnight. Digestion was stopped by acidification with TFA. Bovin Serum Albumin (initial fractionation by heat shock, Sigma) was digested similarly.

**2 LC-MS/MS analysis**

**2.1 Low resolution analysis**

HPLC was performed on a NanoLC-Ultra system (Eksigent). Seven-hundred ng of protein digest were loaded at 7.5 µL/min$^{-1}$ on a precolumn cartridge (stationary phase: C18 PepMap 100, 5 µm; column: 100 µm i.d., 1 cm; Dionex) and desalted with 0.1% HCOOH. After 3 min, the precolumn cartridge was connected to the separating PepMap C18 column (stationary phase: C18 PepMap 100, 3 µm; column: 75 µm i.d., 150 mm; Dionex). Buffers were 0.1% HCOOH in water (A) and 0.1% HCOOH in ACN (B). The peptide separation was achieved with a linear gradient from 5 to 30% B for 60 min at 300 nL/min$^{-1}$. Including the regeneration step at 95% B and the equilibration step at 95% A, one run took 90 min.

Eluted peptides were analysed on-line with a LTQ XL ion trap (Thermo Electron) using a nanoelectrospray interface. Ionization (1.5 kV ionization potential) was performed with liquid junction and a non-coated capillary probe (10 µm i.d.; New Objective). Peptide ions were analysed using Xcalibur 2.0.7 with the following data-dependent acquisition steps: (1) full MS scan (m/z of 300 to 1300, enhanced profile mode) and (2) MS/MS ($qz = 0.25$, activation time = 30 ms, and collision energy = 35%; centroid mode). Step 2 was repeated for the three major ions detected in step 1. Dynamic exclusion was set to 45 s.

**2.2 High resolution (HR) analysis**

HPLC was performed on an Ultimate 3000 LC system (Dionex). Seven-hundred ng of protein digest were loaded at 7.5 µL/min$^{-1}$ on a precolumn cartridge (stationary phase: C18 PepMap 100, 5 µm; column: 300 µm i.d., 5 mm; Dionex) and desalted with 0.08% TFA and 2% ACN. After 3 minutes, the precolumn cartridge was connected to the separating PepMap C18 column (stationary phase: C18 PepMap 100, 3 µm; column: 75 µm i.d., 150 mm; Dionex). Buffers were 0.1% HCOOH, 3% ACN (A) and 0.1% HCOOH, 80% ACN (B). The peptide separation was achieved with a linear gradient from 4 to 36% B for 60 min at 300 nL/min$^{-1}$. Including the regeneration step at 100% B and the equilibration step at 100% A, one run took 90 min.

Eluted peptides were analysed on-line with a LTQ-Orbitrap Discovery (Thermo Electron) using a nanoelectrospray interface. Ionization (1.3 ionization potential) was performed with liquid junction and a non-coated capillary probe (10 μm i.d.; New Objective). Peptide ions were analysed using Xcalibur 2.0.7 with the following data-dependent acquisition steps: (1) FTMS scan on Orbitrap (m/z of 300 to 1300, 15000 resolution, profile mode), (2) MS/MS on the LTQ (qz = 0.25, activation time = 30 ms, and collision energy = 35%; centroid mode). Step 2 was repeated for the two major ions detected in step 1. Dynamic exclusion was set to 90s.

# 3 Protein identification

Database search was performed with the X!Tandem software (version 2010.01.01.4) (http://www.thegpm.org/TANDEM/). Enzymatic cleavage was declared as a trypsin digestion with one possible misscleavage. Cys carboxyamidomethylation and Met oxidation were set to respectively static and possible modifications. Precursor mass precision was set to 2.0 Da for LR and 20 ppm for HR. Fragment mass tolerance was 0.5 Th for both LR and HR data. A refinement search was added with the same parameters except that semi-trypsic peptide and protein N-ter acetylation were also searched. The *Saccharomyces* Genome Database (http://downloads.yeastgenome.org, 5885 entries) was searched together with a contaminant database (trypsin, keratins...). The decoy database comprised the reverse protein sequences of the *Saccharomyces* database. Only peptides with an E-value smaller than 0.1 were reported.

Identified proteins were filtered and sorted by using the XTandem pipeline (http://pappso.inra.fr/bioinfo/xtandempipeline/). Criteria used for protein identification were (1) at least two different peptides identified with an E-value smaller than 0.05, (2) a protein E-value (product of unique peptide E-values) smaller than $10^{-4}$. These criteria led to a False Discovery Rate of 0.3% for peptide and protein identification. To take redundancy into account (i.e. the fact that the same peptide sequence can be found in several proteins), proteins with at least one peptide in common were grouped. Grouped proteins had similar functions. Within each group, proteins with at least one specific peptide relatively to other members of the group were reported as sub-groups.

# 4 Statistical analysis

In order to take into account possible global quantitative variations between LC-MS runs, normalization was performed. For each LC-MS run, the ratio of all peptide values to their value in the chosen reference LC-MS run was computed. Normalization was performed by dividing peptide values by the median value of peptide ratios. Subsequent statistical analyses were performed on $\log_{10}$-transformed normalized data.

BSA peptides showed linear relationships to protein quantity, but the slopes were peptide-specific. These peptide-specific global effects were estimated by a two-way ANOVA on normalized data, where the two factors were peptide and BSA quantity. This enabled the estimation of the specific effect of each BSA peptide, which was removed from normalized data in Figure 3. This allowed the representation of all peptides on the same scale, but had no effect on curve shapes.